\documentclass{article} 
\usepackage{iclr2025_delta,times}

\usepackage{amsmath,amsfonts,bm}

\def\eqref#1{equation~\ref{#1}}

\def\1{\bm{1}}

\DeclareMathAlphabet{\mathsfit}{\encodingdefault}{\sfdefault}{m}{sl}
\SetMathAlphabet{\mathsfit}{bold}{\encodingdefault}{\sfdefault}{bx}{n}

\usepackage{hyperref}
\usepackage{url}
\usepackage{xcolor}
\usepackage{colortbl}
\usepackage{graphicx}
\usepackage{amsmath} 
\usepackage{booktabs}
\usepackage{multirow}
\title{LapLoss: Laplacian Pyramid-based Multiscale loss for Image Translation}

\author{Krish Didwania, Prakhar Arya\textsuperscript{†}, Sanskriti Labroo\textsuperscript{†} \\
Department of Computer Science and Engineering\\
Manipal Institute of Technology \\
Manipal Academy of Higher Education\\
Manipal, Karnataka, India \\
\texttt{\{krishdidwania0674,prakhararya13,sanskritilabroo\}@gmail.com} \\
\And
Ishaan Gakhar\textsuperscript{†}\\
Department of Information and Communication Technology\\
Manipal Institute of Technology \\
Manipal Academy of Higher Education\\
Manipal, Karnataka, India\\
\texttt{ishaangakhar04@gmail.com}
}

\iclrfinalcopy 
\begin{document}


\maketitle

\begingroup
\renewcommand\thefootnote{\textsuperscript{†}}
\footnotetext{Authors have equal contribution to the work.}
\endgroup

\footnotetext{This paper has been accepted at the ICLR 2025 DelTa Workshop.}

\begin{abstract}

Contrast enhancement, a key aspect of image-to-image translation (I2IT), improves visual quality by adjusting intensity differences between pixels. However, many existing methods struggle to preserve fine-grained details, often leading to the loss of low-level features. This paper introduces LapLoss, a novel approach designed for I2IT contrast enhancement, based on the Laplacian pyramid-centric networks, forming the core of our proposed methodology. The proposed approach employs a multiple discriminator architecture, each operating at a different resolution to capture high-level features, in addition to maintaining low-level details and textures under mixed lighting conditions. The proposed methodology computes the loss at multiple scales, balancing reconstruction accuracy and perceptual quality to enhance overall image generation. The distinct blend of the loss calculation at each level of the pyramid, combined with the architecture of the Laplacian pyramid enables LapLoss to exceed contemporary contrast enhancement techniques. This framework achieves state-of-the-art results, consistently performing well across different lighting conditions in the SICE dataset.



\end{abstract}

\section{Introduction}

Image-to-image translation (I2IT) \citep{8100115} is a popular task in the field of Computer Vision, which targets the transfer of images mapped from an input domain to an output domain. It is a critical challenge in modern industries, where the demand for complex and precise image transformations is rapidly growing requiring diverse image transformations, from minor augmentations to major format alterations. As the field advances, I2IT has proven highly effective in tasks such as colourizing grayscale images, image illumination, style transfer, and contrast enhancement, with the latter being a key challenge for improving visual clarity in underexposed or overexposed images, particularly in critical applications like autonomous driving \citep{xia2022imagetoimagetranslationautonomousdriving}, where accurate and realistic visual representations are essential.

Many traditional methods \citep{liu2017unsupervised} \citep{wang2018high} are available for performing translation, but they are computationally heavy, often requiring large inference times due to the complexity of the algorithms involved. This work focuses on tackling contrast enhancement and combatting the existing limitations with a novel approach. To this end, the Laplacian Pyramid \citep{9578037}, a powerful multi-scale image processing approach that decomposes any image into a series of levels that represent distinct low and high-level features and from which the original image can be reconstructed is employed. Initially, it involves the creation of a Gaussian pyramid from the original image through downsampling and at each level, the difference between the original and the smoothened version is calculated to capture finer structural intricacies. This method preserves important features across different scales, enabling models to better handle the structure of an images, which ultimately enhances the quality of image translation.  It addresses the previous challenges through its lightweight architecture \citep{huang2022exposure}, which supports faster inference times while delivering results comparable to other state-of-the-art models (SOTA).

In this work, Generative Adversarial Networks employed (GANs) to enhance the translational networks for I2IT \citep{denton2015laplacian}. The generator produces synthetic images that mimic the ground truth, while the discriminator learns to differentiate between real and generated images \citep{goodfellow2016deep}. This competitive process encourages incremental improvements, refining the generator’s outputs to address subtle changes like variations in saturation and other visual attributes, successfully fooling the discriminator over time. Due to this competitive framework, GANs prove to be highly versatile with their use cases \citep{10445413}; \citep{sauer2023stylegantunlockingpowergans}, and produce highly realistic outputs. However, GANs are susceptible to unstable training and mode collapse \citep{durall2020combatingmodecollapsegan}, where the generator fails to capture the full data distribution and repeatedly produces similar outputs. Additionally, GANs may suffer from vanishing gradients, making optimization difficult. Generative tasks in high-dimensional spaces demand more robust solutions for ensuring efficiency and latent space leads to further instability of training.

Major contributions in this work are summarized as:

\begin{itemize}
    \item Proposed a novel approach that integrates pixel-wise loss with adversarial losses across multiple scales of a Laplacian pyramid, achieving state-of-the-art results.
    \item Performed extensive experiments to analyze the impact of the translational network at each pyramid level, demonstrating its adaptability and generalizability across nine different contrast levels, effectively handling underexposed, overexposed, and mixed-exposure conditions.
    \item Showcased the robustness of the proposed loss function through cross-validation, achieving competitive performance to other frameworks for contrast enhancement.
\end{itemize}

    
    
    
    


\section{Related Works}

\subsection {Traditional Low-Light Enhancement Methods and Contrast Enhancement}

The low-light image enhancement (LLIE) discipline has transitioned significantly over the years. Elementary approaches relied upon histogram equalization \citep{woods2018}, which redistributes pixel intensities across the entire range. Further contrast enhancement in each tile and modifying the histogram equalization process \citep{reza2004, tian2017}. Despite their utility, the persistent shortcomings of histogram-based methods in achieving accurate colour restoration—particularly under non-linear illumination distortions—] motivated the adoption of gamma correction \citep{huang2012, huang2013} by applying power-law transformation compensated for the non-linear response of display devices and for enhancing details in dark or bright regions of an image. However, it does not address issues like noise or colour distortion, which are common in low-light conditions.

Deep learning has improved LLIE by addressing colour shifts, noise, and uneven illumination. RetinexNet \citep{Chen2018Retinex, wang2019} and Bread \citep{guo2023lowlight} tackle noise but introduce colour distortion. SNR-Aware \citep{xu2022} tries to tackle noise using transformers but struggles with extreme lighting. Recent approaches, including DRBN \citep{yang2020} and self-supervised methods \citep{li2021}, have improved noise suppression and adaptive illumination adjustment. Another direction of using Diffusion models led to innovations including DDPMs \citep{ho2020}, PyDiff \citep{zhou2023}, and Diff-Retinex \citep{yi2023} which improved noise handling but suffered from inefficiency and over-exposure. KinD \citep{zhang2019kindling} introduces noise-aware priors, yet the high costs and distortions remain.

Generative models, especially GANs and Cycle GANs, have shown incredible prowess in contrast enhancement across domains. In the medical domain, GANs have been applied to generate contrast-enhanced MRI and CT images \citep{Cheng2024}. For example, CGAN-based models yield notable SSIM values for the synthetic T1-weighted brain MRI images \citep{gen1}; the deep learning frameworks for sCECT enhance mediastinal lymph nodes lesion conspicuity and contrast-to-noise ratios \citep{gen2}. Nonetheless, there are significant limitations for generative models \citep{Skandarani2023}, which limit their wider applicability. For instance, in medical imaging, their performance mostly relies on homogeneous data sets and often shows less generalization \citep{gen2}. Moreover, their heavy computational nature postpones real-time deployment \citep{gen1} \citep{10172181} hindering practical capabilities.

\subsection{Laplacian Pyramids in I2IT}

Pyramid decomposition, originally used for multi-resolution analysis like DWT \citep{1095851}, has been widely adopted in machine learning. Techniques like using convolutional networks in a Laplacian pyramid for image generation \citep{denton2015laplacian} and SPD for textures \citep{thakur2015steerable} have advanced this approach. Our work builds on this by employing a Laplacian pyramid to decompose images, enabling better high-resolution detail restoration. Parallely, The fastFF2FFPE method \citep{10.1007/978-3-031-16434-7_40}, which uses a Laplacian Pyramid to decompose FF histo-pathological images into low- and high-frequency components for efficient FFPE-style translation, offers notable computational advantages, including faster inference and lower memory usage compared to methods like vFFPE and AI-FFPE. However, it is quantitative results and perceptual quality remain comparable without surpassing existing approaches, limiting its appeal for high-fidelity applications and high-frequency details.

A recent development of the Laplacian Pyramid Translation Network (LPTN) \citep{9578037} is a scalable deep learning framework for image enhancement tasks such as low-light enhancement. LPTN leverages a Laplacian Pyramid to decompose images into low-frequency global features and high-frequency components independently processed by lightweight neural networks, enhancing global and local features while minimizing computational costs. The Laplacian Pyramid Super-Resolution Network (LapSRN) \citep{LapSRN} progressively reconstructs high-resolution images by predicting residuals in a coarse-to-fine manner using a Laplacian pyramid framework. Despite the improvements made, LPTN suffers from limitations in addressing mixed exposures, where it fails to focus on localized behaviour and does not adapt to changes in dynamic lighting \citep{8796366}, curtailing its ability to perform at an optimal level on challenging datasets. The research by \citep{rathore2025hipyrnethypernetguidedfeaturepyramid} highlights the effectiveness of the LPTN architecture when adapted to process both underexposed and overexposed images, albeit with a significant computational cost.

\section{Proposed Methodology}

\subsection{Architecture}

\begin{figure}[t]
    \centering
        \caption{Schematic overview of the LapGSR model \citep{kasliwal2024lapgsr} employed for multi-level adversarial processing (Fig. \ref{fig:Generator_Methodl}). Each branch helps to extract features to finally non-parametrically reconstruct the output image. The instance in the figure is taken from the test set of SICE V1 dataset \citep{zheng2024lowlightimagevideoenhancement}. }
    \includegraphics[width=\textwidth, height=3.5in]{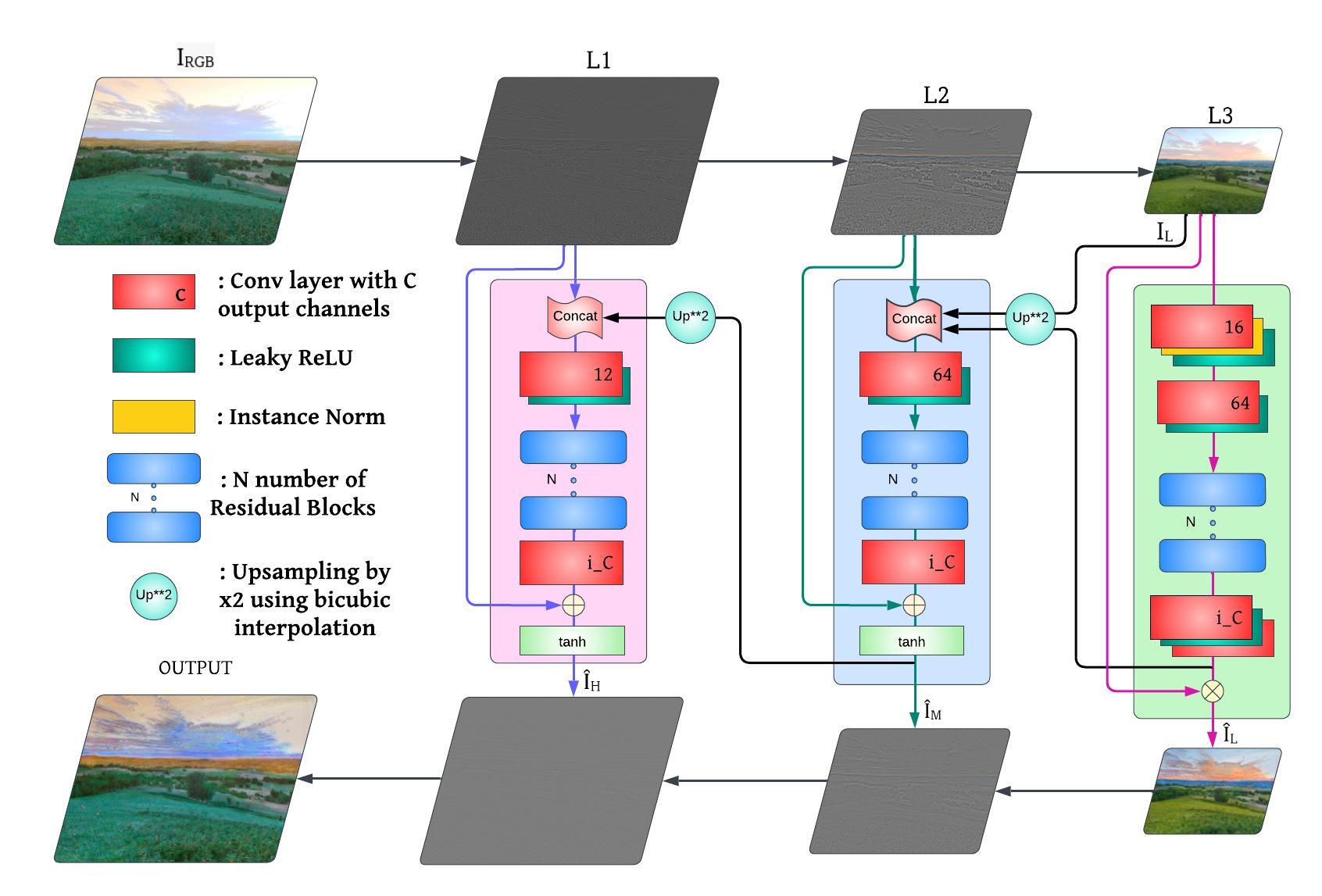}

    \label{fig:lapgsr_model}
\end{figure}

This section describes the architecture employed to generate the Laplacian pyramids of the contrast-enhanced image. LapGSR \citep{kasliwal2024lapgsr}, a lightweight model designed for Guided Super Resolution is employed for this task. With some minor changes to the model, it is configured for single-image processing and reconstruction of Laplacian pyramids. As evident in Fig. \ref{fig:lapgsr_model}, LapGSR merges features from various levels using residual blocks in 3 different branches to reconstruct a Laplacian pyramid. Each branch of the model extracts features from the Laplacian pyramid and reconstructs the corresponding pyramid layer. The number of residual blocks in each layer is represented as $N_{Top}, N_{Middle}$ and $N_{Low}$. The ideal configuration of these residual blocks is decided by exhaustive experimentation. This pyramid is responsible for the final non-parametric reconstruction of the output image by an inverse Laplacian operation. Here, we modify the pipeline to preserve each layer of the pyramid and use it to compute loss against the Laplacian pyramid of the Ground Truth, as represented in Fig. \ref{fig:Generator_Methodl}. A more detailed explanation of the architecture is included in the appendix \ref{apendix}.


Each discriminator, corresponding to a level of the translational network's output, contains 4 residual blocks for level 0, 4 residual blocks for level 1, and 3 residual blocks for level 2. Its architecture also includes instance normalization \cite{Ulyanov2016InstanceNT} and Leaky ReLU \cite{maas2013rectifier}.

\subsection{Loss Function}
\label{section: loss_method}
In the proposed method, a distinct discriminator is introduced at each level of the Laplacian pyramid, where it is trained to identify whether the images at that level of the reconstructed output from the generator are real or fake. Each level contributes to the final loss, as a result of a weighted average of the losses from different levels. Adversarial loss was combined with a pixel-wise  \(\mathcal{L}_{\text{MSE}}\) Loss at various scales to demonstrate results and robustness across datasets.


\textbf{Mean Squared Error Loss (MSELoss)} \citep{goodfellow2016deep} is a fundamental loss function commonly used in image-based tasks. It calculates the pixel-wise difference between the squares of the ground truth and the predicted image, averaged across all pixels. This loss ensures that pixel-wise accuracy between the output and the target images is preserved by penalizing larger deviations more severely to maintain structure.


\textbf{Adversarial Loss} is critical to generating realistic images through a Generative Adversarial Network (GAN) framework, it helps both the generator and the discriminator learn from each other.
We employ the Least-Square Generative Adversarial Network (LSGAN) loss \citep{8237566} along with pixel-wise MSE to enhance image fidelity by preserving critical spatial attributes. The LSGAN loss for the discriminator and generator is expressed as:
\begin{equation}
\mathcal{L}_{\text{D}} = \frac{1}{2} \mathbb{E}_{x_{\text{real}}}[(D(x_{\text{real}}) - 1)^2] + \frac{1}{2} \mathbb{E}_{x_{\text{fake}}}[D(x_{\text{fake}})^2]
\end{equation}

\begin{equation}
\mathcal{L}_{\text{GAN}} = \frac{1}{2} \mathbb{E}_{x_{\text{fake}}}[(D(x_{\text{fake}}) - 1)^2]
\end{equation}

\begin{figure}[t]
    \centering
        \caption{Schematic overview of the multi-scale GAN paradigm. The affine effect is only applied for visual purposes and is not a preprocessing step. It shows the decomposed pyramid of the predicted image and the ground truth across 3 levels. In the final loss, $\lambda_{1}$, $\lambda_{2}$, and $\lambda_{3}$ are the hyperparameters to weight the level-wise loss explained in Section \ref{section: loss_method}.}
    \includegraphics[width=4.5in, height=3in]{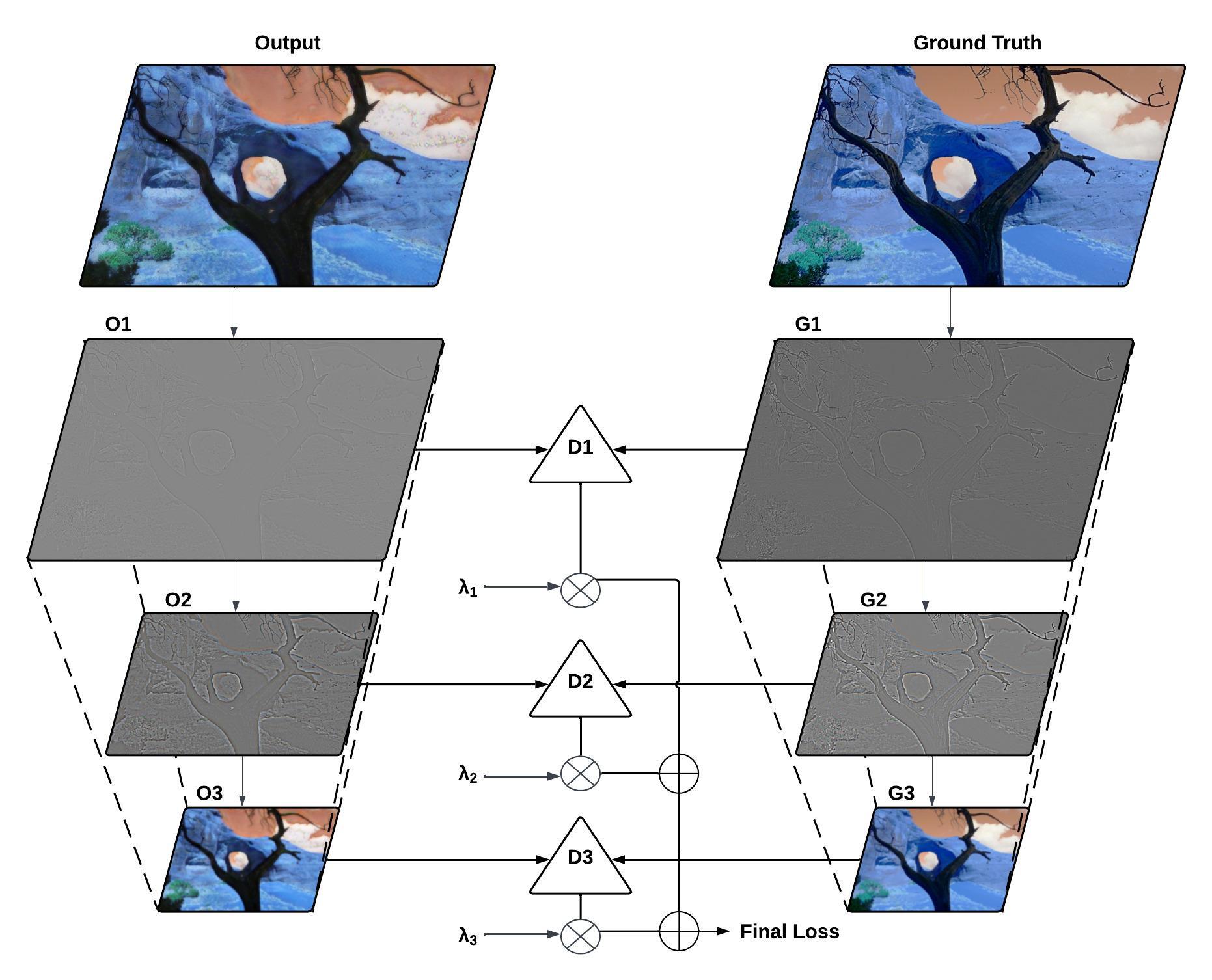}
    \label{fig:Generator_Methodl}
\end{figure}

As illustrated in Fig. \ref{fig:Generator_Methodl}, both the output and ground truth images are decomposed into hierarchical Laplacian pyramid representations. The loss at each level is computed between corresponding pyramid layers, with \( D_1, D_2, D_3 \) discriminators. These losses are weighted using \( \lambda_1, \lambda_2, \lambda_3 \), allowing finer control over how much each scale contributes to the final optimization. This structure aligns well with the LapGSR method, which employs multi-scale, lightweight translational networks where higher-level branches rely on outputs from lower-level branches. Specifically, these values are weighted and summed to enable the generator network to optimize both the precision of the generated images (measured by MSELoss) and its ability to deceive the discriminator (indicated by the adversarial loss). This balances the loss at each level, which is later scaled using a weighting parameter \textit{w}. Therefore, the final loss for the generator is defined as:

\begin{equation}
\label{finaloss}
\mathcal{L}_{\text{total}} = \sum_{i=0}^{N} \lambda_i (\mathcal{L}_{\text{GAN}}^i + \textit{w}  
 \mathcal{L}_{\text{MSE}}^i)
\end{equation}

where \( N \) represents the number of pyramid levels, \( \lambda \) is the weight assigned to each level, \( \mathcal{L}_{\mathrm{MSE}}^i \) denotes the pixel-wise MSELoss at level \( i \), and \( \mathcal{L}_{\mathrm{GAN}}^i \) represents the adversarial loss for that level. Extensive experimentation was conducted with hyperparameters as discussed in Section \ref{section-exp}.



\section{Experiments}
\label{section-exp}

\subsection{Datasets}
The dataset used in this work is the publicly available \textbf{SICE dataset} \citep{cai2018learning}, which comprises 589 high-resolution multi-exposure sequences with a total of 4,413 images.  Each sample in the dataset contains either 7 or 9 different contrast levels of the same scene. For this study, we utilized the \textbf{SICE V2} dataset, which includes 1,458 images derived from the 229 unique (out of the 589) samples for training. 

For testing, images from the \textbf{SICE V1} dataset were used, selecting the -1EV (Exposure value) image as the low-light input for underexposure and the +1EV image for overexposure, creating two separate test sets, as per the testing indices provided with the dataset. Additionally, we utilized the \textbf{SICEGrad} and \textbf{SICEMix} \citep{zheng2024lowlightimagevideoenhancement} datasets, each containing 529 unique images. These datasets were employed for testing, as they include all possible contrast variations within a single image, replicating the training set. The SICEGrad dataset arranges contrast in increasing or decreasing strips, while the Mix dataset has unordered contrast variations. 

\subsection{Hyperparameter Tuning}

The images used for training were resized to a shape of \( 608 \times 896 \), and Vertical Flip, Horizontal Flip, and Shift Scale Rotate augmentations were employed to avoid overfitting of the generator.  

Through extensive experimentation, we determined that the optimal ratio of adversarial to reconstruction weighting to be  
\textit{w}=\( 4.5 \times 10^3 \) which effectively stabilized the translation network. Additionally, we found that a learning rate of \( 10^{-3} \) was optimal for achieving stable convergence, ensuring robust results.


\begin{table}[ht]
\caption{Impact of Level-wise weights on performance across all subsets and datasets. The values show the levels included for loss calculation and their respective weights. The metrics in this table are given as PSNR/SSIM, with higher values indicating better performance.}
\centering
\begin{tabular}{cc|cccc} \hline
Levels      & Weights       & Overexposure  & Underexposure  & SICEGrad  & SICEMix   \\ \hline
{[0]}                 & {[1]}                  & 16.54/0.767  & {16.97}/0.726 & 16.14/0.691 & 16.09/0.680 \\ 
{[1]}                 & {[1]}                  & 18.68/\textbf{0.774}  & 17.74/\textbf{0.748} & {16.49}/\textbf{0.699} & {16.31}/\textbf{0.726} \\ 
{[2]}                 & {[1]}                  & \textbf{18.79}/0.530 & \textbf{18.94}/0.625 & \textbf{17.00}/0.648 & \textbf{16.79}/0.635 \\ \hline
{[0,1,2]}             & {[4/7, 2/7, 1/7]}      & 19.91/0.714 & 18.79/0.751 & 16.74/0.678 & 16.63/0.668 \\ 
{[0,1,2]}             & {[1/7, 2/7, 4/7]}      & 19.71/0.691 & 18.87/0.746 & 16.72/\textbf{0.683} & 16.57/0.671 \\ 
{[0,1,2]}             & {[1/3, 1/3, 1/3]}      & \textbf{20.33}/\textbf{0.745} & \textbf{18.96}/\textbf{0.754} & \textbf{16.76}/0.671 & \textbf{16.64}/\textbf{0.681} \\ \hline
\end{tabular}

\label{tab:levels_weights_metrics}
\end{table}


The interplay between pyramid-level weighting and performance is shown in Table \ref{tab:levels_weights_metrics}. Training with only a single-level loss resulted in higher pyramid levels performing better on PSNR, as these focus on fine-grained details important for pixel-wise accuracy. However, SSIM peaked at intermediate levels, since mid-frequency features control structural coherence. In this setup, training with a single-level loss means that only one discriminator was active at a specific pyramid level during GAN training, with both reconstruction and adversarial losses computed exclusively at that level, resulting in $N$=1 for Equation \ref{finaloss}. Additionally, $i$ represents the corresponding level, as shown in the first three rows of \ref{tab:levels_weights_metrics}.


Weighting schemes like $\frac{1}{7}, \frac{2}{7}, \frac{4}{7}$, which emphasize finer levels, degraded performance because they disproportionately prioritize high-frequency details at the cost of mid-frequency textures and global illumination corrections.This shows that multi-level integration is indeed necessary and the equal weights  of $\frac{1}{3}$ per level optimally allow balanced contributions from all levels, achieving state-of-the-art metrics.

Thorough experimentation was conducted to evaluate the various GAN variants, revealing that LSGAN outperformed WGAN (Wasserstein GAN)\cite{wgan}, WGAN-Softplus \cite{ding2020subsampling}, and HingeGAN\cite{hinge}. Its stable training dynamics and features, such as preventing gradient vanishing, made LSGAN the optimal choice.We employed SOAP (Shampoo with Adam in the Preconditioner's eigenbasis) \citep{vyas2025soapimprovingstabilizingshampoo} for optimizing the generator and AdamW \cite{adam} for each discriminator in the GAN training process.

Furthermore, the number of residual blocks in LapGSR was systematically adjusted after testing configurations ranging from 3 to 5 blocks for each translational network, with trainable parameters increasing from 694K to 1.13M, as detailed in the apendix \ref{apendix}. The explanation and metrics related to these configurations are provided in the ablation section. Ultimately, we proceeded with LSGAN and the 3, 3, 3 lightweight framework for further experiments, comparing our results with those of other approaches.

Structural Similarity Index Measure (SSIM) \citep{1284395} and Peak Signal-to-Noise Ratio (PSNR) \citep{4560230} were used as the metrics of evaluation due to their complementary strengths in measuring the quality of the image. PSNR is pixel-level reconstruction accuracy, quantifying noise and distortion. SSIM evaluates perceptual quality based on luminance, contrast, and structural similarity, making it closer to human visual perception. 





Through extensive hyperparameter tuning, our architecture achieved state-of-the-art results on the SICE dataset. This systematic optimization highlights the importance of customized hyperparameter strategies in improving low-light and high-light image enhancement, showcasing the effectiveness of our Laplacian pyramid-based approach.

\section{Results}


\begin{figure}[t]
    \centering
    \caption{Input and output taken for various samples across all datasets. The images in the 1st row are taken from the Overexposure set, 2nd row is taken from Underexposure, 3rd are taken from SICEMix and 4th are taken from SICEGrad. All images are from the test sets.}
    \includegraphics[width=5in , keepaspectratio]{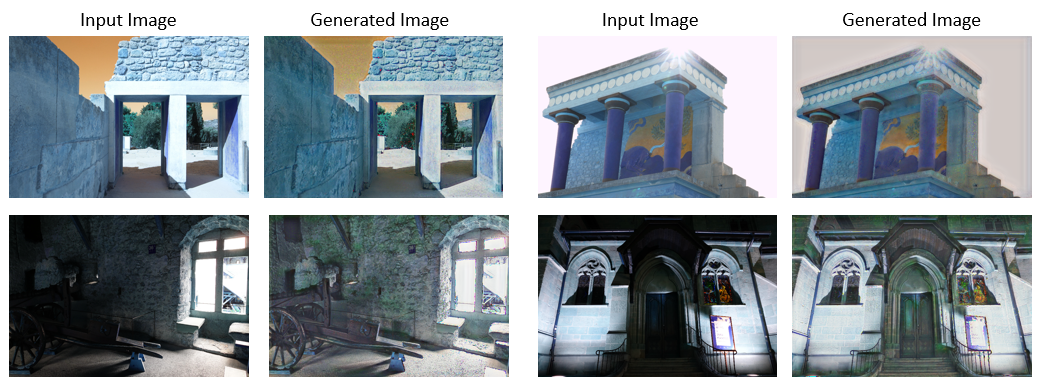} \\
    \includegraphics[width=5in, keepaspectratio]{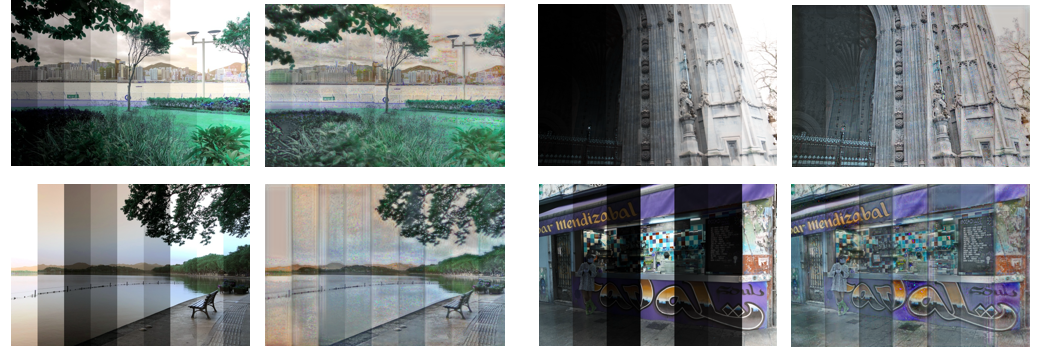} 
    \label{fig:combined}
\end{figure}


The proposed method, LapLoss, effectively mitigates illumination inconsistencies such as non-uniform noise and abrupt luminance transitions. Exhaustive experiments conducted on the SICE mixed-illumination testing sets (SICEGrad and SICEMix) demonstrate our method's ability to balance striated darkness—alternating bands of underexposed and well-lit regions found in cloud-shadowed landscapes or unevenly lit interiors. As shown in Fig. \ref{fig:combined}, the framework seamlessly corrects these inconsistencies, producing outputs closely aligned with ground truth images.


Trained on images with nine contrast levels, ranging from extreme underexposure to overexposure, it effectively corrects contrast across diverse lighting conditions. As demonstrated in Fig. \ref{fig:combined}, the framework enhances both brightly lit and dimly lit conditions. Integrating Laploss with LapGSR ensures the preservation of textures, colours, and details across varying illumination, thereby demonstrating excellent generalizability.








\begin{table}[h]
    \centering
    \caption{Comparing our results on SICE test sets against other models. 
    The best, second-best, and third-best metrics per column are highlighted 
    in \textcolor{blue!80}{blue}, \textcolor{orange!80}{orange}, and 
    \textcolor{red!80}{red}, respectively. \textsuperscript{\textdagger}Experiments conducted by authors.}
    
    \resizebox{\linewidth}{!}{%
\begin{tabular}{lcccccc}
\toprule
\multirow{2}{*}{Method} & \multicolumn{2}{c}{Underexposure} & \multicolumn{2}{c}{Overexposure} & \multicolumn{2}{c}{Average} \\
\cmidrule(lr){2-3} \cmidrule(lr){4-5} \cmidrule(lr){6-7}
& PSNR $\uparrow$ & SSIM $\uparrow$ & PSNR $\uparrow$ & SSIM $\uparrow$ & PSNR $\uparrow$ & SSIM $\uparrow$ \\
\midrule
LCDPNet~\citep{wang2022local}           & 17.45 & 0.562 & 17.04 & 0.646 & 17.25 & 0.604 \\
DRBN~\citep{yang2020}              & 17.96 & 0.677 & 17.33 & 0.683 & 17.65 & 0.680 \\
DRBN+ERL~\citep{Huang_2023_CVPR}          & 18.09 & 0.674 & 17.93 & 0.686 & 18.01 & 0.680 \\
DRBN-ERL+ENC~\citep{Huang_2023_CVPR}      & 22.06 & 0.705 & 19.50 & 0.721 & 20.78 & 0.713 \\
ELCNet~\citep{huang2017arbitrary}            & 22.05 & 0.689 & 19.25 & 0.687 & 20.65 & 0.686 \\
IAT~\citep{cui2022you}               & 21.41 & 0.660 & \textcolor{blue!80}{22.29} & 0.681 & \textcolor{orange!80}{21.85} & 0.671 \\
ELCNet+ERL~\citep{Huang_2023_CVPR}        & \textcolor{red!80}{22.14} & 0.691 & 19.47 & 0.692 & 20.81 & 0.695 \\
FECNet~\citep{huang2019}            & 22.01 & 0.674 & 19.91 & 0.696 & 20.96 & 0.685 \\
FECNet+ERL~\citep{Huang_2023_CVPR}        & \textcolor{orange!80}{22.35} & 0.667 & 20.10 & 0.689 & \textcolor{red!80}{21.22} & 0.678 \\
U-EGformer~\citep{Adhikarla2024}                 & 21.63 & \textcolor{red!80}{0.711} & 19.74 & 0.705 & 20.69 & \textcolor{red!80}{0.707} \\
U-EGformereaf~\citep{Adhikarla2024}               & \textcolor{blue!80}{22.98} & \textcolor{orange!80}{0.719} & \textcolor{orange!80}{21.84} & \textcolor{orange!80}{0.710} & \textcolor{blue!80}{22.41} & \textcolor{orange!80}{0.717} \\
\bottomrule

{LPTN+LapLoss}\textsuperscript{\textdagger} & 18.94 & 0.653& 20.26& 0.698& 19.60&0.676 \\

{LapGSR\textsuperscript{\textdagger}} & 19.33	&0.662&20.62&	\textcolor{red!80}{0.708}& 19.97 & 0.685 \\

{LapGSR+LapLoss}\textsuperscript{\textdagger} & 19.42 & \textcolor{blue!80}{0.732}& \textcolor{red!80}{21.32}& \textcolor{blue!80}{0.766}& 20.37 & \textcolor{blue!80}{0.749} \\

\bottomrule
\label{sice-overunder}
\end{tabular}
}
\end{table}

\begin{table}[h]
\centering
\caption{Comparison of results of the proposed method on SICEGrad and Mix sets against other models. The best, second-best, and third-best metrics per column are highlighted in \textcolor{blue!80}{blue}, \textcolor{orange!80}{orange}, and \textcolor{red!80}{red}, respectively. \textsuperscript{\textdagger}Experiments conducted by authors.}
\begin{tabular}{lcccc}
\toprule
\multirow{2}{*}{Method} & \multicolumn{2}{c}{SICEGrad} & \multicolumn{2}{c}{SICEMix} \\
        \cmidrule(lr){2-3} \cmidrule(lr){4-5}
        & PSNR $\uparrow$ & SSIM $\uparrow$ & PSNR $\uparrow$ & SSIM $\uparrow$ \\
        \midrule
RetinexNet \citep{Wei2018} & {12.40} & {0.606} & {12.45} & {0.619} \\
ZeroDCE \citep{Guo_2020_CVPR} & {12.43} & {0.633} & {12.48} &{ 0.644} \\
RAUS \citep{Zhang2021} &{ 0.86} & {0.493} & {0.86} & {0.494} \\
SGZ \citep{Zheng2021} & 10.86 & 0.607 & 10.99 & 0.621 \\
LLFlow \citep{Wang2021} & 12.74 & 0.617 & 12.74 & 0.617 \\
URetinexNet \citep{Wu2022} & 10.90 & 0.600 & 10.89 & 0.610 \\
SCI \citep{Ma2022} & 8.64 & 0.529 & 8.56 & 0.532 \\
KinD \citep{Zhang2021} & 12.99 & 0.656 & 13.14 & \textcolor{orange!80}{0.668} \\
KinD++ \citep{Zhang2021} & 13.20 & 0.657 & 13.24 &  \textcolor{red!80}{0.666} \\
U-EGformer \citep{Adhikarla2024} & 13.27 & 0.643 & 14.24 & 0.652 \\
U-EGformer$^{\dagger}$ \citep{Adhikarla2024} & 14.72 & \textcolor{red!80}{0.665} & 15.10 & \textcolor{blue!80}{0.670} \\
\midrule
LPTN+LapLoss\textsuperscript{\textdagger}&  \textcolor{orange!80}{17.32} &	 \textcolor{orange!80}{0.657} & \textcolor{red!80}{16.67} &	0.624 \\
LapGSR\textsuperscript{\textdagger} & \textcolor{red!80}{17.27}&0.629&  \textcolor{orange!80}{16.71}	&0.623 \\

LapGSR+LapLoss\textsuperscript{\textdagger} & \textcolor{blue!80}{17.33}	& \textcolor{blue!80}{0.657} & \textcolor{blue!80}{17.13} &	0.648 \\
\bottomrule

\label{sice-mixgrad}
\end{tabular}
\end{table}


As shown in Table \ref{sice-overunder}, our proposed model achieves the highest SSIM across all lightweight models in recent years. While it does not produce state-of-the-art PSNR on the overexposure and underexposure test sets, it outperforms other methods in SSIM, demonstrating superior image structure preservation. This improvement over previous models highlights its ability to maintain texture consistency and natural luminance gradients, which are crucial for human perception. Results of LapGSR outdo LPTN with Laploss, which are further outperformed by LapGSR and Laploss. The significant increase of \textbf{10\%} in a few metrics between LapGSR and LapGSR with LapLoss thereby validates our methodology.

Shifting our focus to Table \ref{sice-mixgrad}, we aim to establish the robustness of our method and achieve notable enhancements in both PSNR and SSIM across the SICEGrad and SICEMix at only \textbf{694K} trainable parameters. An increase of \textbf{30\%} compared to the previous SOTA is observed in the PSNR on the SICEGrad dataset. Similarly, a \textbf{20\%} improvement is noted in PSNR on the SICEMix dataset. Along with this, a steady increase in SSIM is observed across all datasets. Thus, the proposed loss propels us to excel at scenarios where abrupt lighting transitions or spatially varying exposures challenge conventional methods. A notable difference in performance is noted between LapGSR and LapGSR with LapLoss, validating our proposed methodology.



\section{Conclusion and Future Works}
\vspace{-0.2cm}
In this work, we introduced \textbf{LapLoss}, a novel approach which finds applicability to a variety of tasks in I2IT, utilizing Laplacian Pyramid Translational Networks with varying architectures. Our findings demonstrate the adaptability and generalizability of the proposed approach in restoring images across diverse contrast levels. Additionally, an in-depth analysis of the contribution of each level to the overall framework's performance is presented, highlighting the significance of multi-level processing in achieving superior results.

For future work, we aim to extend our experiments with a broader range of generator loss functions and pixel-wise loss functions, to enhance robustness and accuracy.
We believe that the insights gained from this study can be applied to other domains, such as super-resolution, debanding, and denoising to pave the way for a more unified approach to solving various image restoration problems, leveraging the power of LPTNs and advanced loss functions to deliver cutting-edge results.

\section*{Acknowledgments}  
The authors would like to thank the Research Society Manipal, a student-run organization at Manipal Institute of Technology, Manipal, for providing the necessary resources for this work. We also express our gratitude to Aditya Kasliwal and Pratinav Seth for their valuable contributions in refining the quality of the final manuscript.

\bibliography{iclr2025_delta}
\bibliographystyle{iclr2025_delta}

\newpage
\appendix
\section{Appendix}
\label{apendix}
In this study, we detail the exact architecture, effect and configuration of the residual blocks. Additionally, we display results with various Adversarial Losses, as well as their effects across datasets.

\subsection{Residual Blocks}

A detailed explanation of the architecture used for experiments as shown in Fig. \ref{fig:lapgsr_model} is given below:\\
\textbf{Lower Transformation Branch (LTB):} \\
The LTB extracts fundamental features such as luminance, texture, and illumination from lower most level of the pyramid. It starts with a convolutional layer, followed by instance normalization and a leaky ReLU activation to prevent vanishing gradients. After this, another convolutional layer with leaky ReLU is applied, followed by several residual blocks consisting of convolutional layers with skip connections. The final feature map, denoted as $\hat{I}_L$, is the output of the LTB. This feature map is upsampled and combined with the second-to-last layer ($L_2$) of the branch before being passed to the Middle Transformation Branch (MTB). The output of the residual blocks is multiplied with $L_1$, yielding the final product of the LTB.

\textbf{Middle Transformation Branch (MTB):} \\
The MTB serves to bridge the low-level features from the LTB and the high-level abstractions for subsequent tasks. It begins with a convolutional layer followed by leaky ReLU activation to extract features while avoiding vanishing gradients. The MTB contains several residual blocks, followed by a final convolutional layer. The feature maps from the LTB and Laplacian pyramid ($L_2$) are fused and passed through a tanh activation, yielding the intermediate representation, $\hat{I}_M$. This output is upsampled by 2x and concatenated with $L_3$, before being passed into the High Transformation Branch (HTB).

\textbf{High Transformation Branch (HTB):} \\
The HTB is the final stage of the transformation pipeline, specializing in synthesizing the contrast enhanced output image. It receives the 2x upsampled output from the MTB, $\hat{I}_M$, and processes it through a convolutional layer, followed by leaky ReLU activation. The residual blocks refine the upsampled features, and a final convolutional layer consolidates them into a corrected contrast feature map. This output is then added to $L_1$ and passed through a tanh activation to generate the top layer, $\hat{I}_H$, of the translated pyramid. This layer combines detailed texture and abstract features for the final visual output.

Through systematic evaluation as shown in Table \ref{tab:res_block_exps}, employing 5 residual  for the lowest level branch ($N{Low}$), 5 for the intermediate level ($N{\text{Mid}}$), and 5 for the top level ($N{\text{Top}}$) in the network for balancing the feature depth and computational efficiency. Although the configuration of 5,5,5 in low, middle and top branches achieved the best performance in terms of metrics, we present results using the 3,3,3 configuration while comparing against other models, as the difference in metrics was not substantial despite the increased framework complexity. We also observed that the number of residual blocks in the low-frequency component played a crucial role in the overall quality of the generated images, as it serves as the foundation for translation in our interconnected network.

\begin{table}[ht]
\centering
\begin{tabular}{cccccccc}
\hline
\textbf{Params} & \textbf{Low} & \textbf{High} & \textbf{Top} & \multicolumn{2}{c}{\textbf{Over}} & \multicolumn{2}{c}{\textbf{Under}} \\
\cline{5-6} \cline{7-8}
 & & & & \textbf{PSNR} & \textbf{SSIM} & \textbf{PSNR} & \textbf{SSIM} \\
\hline

768k & 4 & 3 & 3 & 21.08 & 0.763 & 19.71 & 0.741 \\
842k & 5 & 3 & 3 & 21.21 & 0.742 & 19.31 & 0.7206 \\
916k & 5 & 4 & 3 & 21.12 & 0.773 & 19.46 & 0.7239 \\
990k & 5 & 5 & 3 & 21.06 & 0.771 & 18.81 & 0.7199 \\
1.06M & 5 & 5 & 4 & 21.09 & 0.766 & 18.91 & 0.7049 \\
1.14M & 5 & 5 & 5 & 21.32 & 0.765 & 19.42 & 0.7324 \\
\hline
\end{tabular}
\caption{Test set results for underexposure and overexposure sets.}
\label{tab:res_block_exps}
\end{table}

\begin{table}[h]
\centering
\begin{tabular}{cccccccc}
\hline
\textbf{Params} & \textbf{Low} & \textbf{High} & \textbf{Top} & \multicolumn{2}{c}{\textbf{Grad}} & \multicolumn{2}{c}{\textbf{Mix}} \\
\cline{5-6} \cline{7-8}
 & & & & \textbf{PSNR} & \textbf{SSIM} & \textbf{PSNR} & \textbf{SSIM} \\
\hline

768k & 4 & 3 & 3 & 17.25 & 0.658 & 16.94 & 0.644 \\
842k & 5 & 3 & 3 & 17.33 & 0.657 & 17.12 & 0.647\\
916k & 5 & 4 & 3 & 17.28 & 0.654 & 17.03 & 0.642\\
990k& 5 & 5 & 3 & 17.26 & 0.656 & 17.03 & 0.647 \\
1.06M & 5 & 5 & 4 & 17.05 & 0.642 & 17.33 & 0.658 \\
1.13M & 5 & 5 & 5 & 17.28 & 0.659 & 17.05 & 0.648 \\
\hline
\end{tabular}
\caption{Test set results with two metrics for Grad and Mix datasets.}
\end{table}

Another observation we made is that increasing the number of residual blocks in the intermediate layers can lead to a decrease in metrics. This could be due to the fact that, in this particular case of I2IT, it is crucial to preserve the frequency components. In contrast, enhancement techniques like the intermediate level of a Laplacian pyramid effectively manage this by focusing on the finer details of the frequency spectrum. The intermediate level of a pyramid captures crucial frequency components that help maintain the balance between high-level features and low-level details, which can lead to improved performance in tasks that require fine-grained information preservation.

\subsection{Adversarial Losses}

\begin{table}[h]
\caption{Performance metrics (PSNR/SSIM) for different GAN types across four testing sets. Best metrics are highlighted.}
\label{tab:gan_metrics}
\centering
\begin{tabular}{lcccc}
\hline
\textbf{Loss Type} & \textbf{Overexposure} & \textbf{Underexposure} & \textbf{SICEMix} & \textbf{SICEGrad} \\
\hline
LSGAN         & \textbf{20.54}/\textbf{0.75}   & \textbf{18.88}/\textbf{0.75}    & 16.79/0.67    & 16.66/0.66 \\
WGAN          & 20.00/0.73                   & 18.78/0.73            & 16.75/\textbf{0.68}    & 16.62/\textbf{0.67} \\
WGAN\_SOFT+   & 20.13/0.75                   & 18.72/0.72            & \textbf{16.80}/0.67    & 16.61/0.66 \\
HINGE         & 19.02/0.68                   & 19.65/0.72     & 16.80/0.65    & \textbf{16.64}/0.64 \\
\hline
\end{tabular}

\end{table}

Our proposed loss function for Laplacian pyramid-centric generators primarily focuses on adversarial losses at each level. To evaluate its effectiveness, we experimented with various GAN loss functions, as summarized in Table \ref{tab:gan_metrics}. The results indicate that most loss functions performed similarly, with minor variations in performance metrics.

Based on our experiments, LSGAN consistently produced the highest-quality images, achieving the best average metrics across all four testing sets. As shown in Table \ref{tab:gan_metrics}, LSGAN outperformed other GAN loss functions, particularly excelling in the Overexposure test set. This highlights its robustness in handling bright regions and its overall effectiveness in image enhancement tasks.

\end{document}